# Phase Composition of AlTiNbMoV, AlTiNbTaZr and AlTiNbMoCr Refractory Complex Concentrated Alloys: A Correlation of Predictions and Experiment

JIŘÍ KOZLÍK, FRANTIŠEK LUKÁČ, MARIANO CASAS LUNA, KRISTIÁN ŠALATA, JOSEF STRÁSKÝ, JOZEF VESELÝ, ELIŠKA JAČA, and TOMÁŠ CHRÁSKA

Designing complex concentrated alloys (CCA), also known as high entropy alloys (HEA), requires reliable and accessible thermodynamic predictions due to vast space of possible compositions. Numerous semiempirical parameters have been developed for phase predictions over the years. However, in this paper we show that none of these parameters is a robust indicator of phase content in various refractory CCA. CALPHAD proved to be a more powerful tool for phase predictions, however, the predictions face several limitations. AlTiNbMoV, AlTiNbTaZr and AlTiNbMoCr alloys were prepared using blended elemental powder metallurgy. Their phase and chemical composition were investigated by the means of scanning electron microscopy, energy-dispersive X-ray spectroscopy and X-ray diffraction. Apart from the minor contamination phases ($Al_2O_3$ and Ti(C,N,O)), AlTiNbMoV and AlTiNbMoCr exhibited single-phase solid solution microstructure at the homogenization temperature of 1400 °C, while $Al_3Zr_5$ based intermetallics were present in the AlTiNbTaZr alloy. None of the simple semiempirical parameter was able to predict phase content correctly in all three alloys. Predictions by CALPHAD (TCHEA4 database) were able to predict the phases with limited accuracy only. Critical limitation of the TCHEA4 database is that only binary and ternary phase diagrams are assessed and some more complex phases cannot be predicted.

## I. INTRODUCTION

COMPLEX concentrated alloys (CCAs) are a novel class of material, consisting of multiple elements where none of them is major or dominant. Among them, the so-called refractory complex concentrated alloys (RCCAs) are those containing mostly refractory elements (Ti, V, Cr, Zr, Nb, Ta, Hf, W, Mo,...) with a tendency to form a BCC solid solution phase.[1] As the name suggests, these alloys are candidates for high-temperature applications, possibly surpassing the currently used nickel-based superalloys. However, as a novel alloy class, they suffer from the lack of experimental data documenting their properties. Furthermore, successful design of CCAs requires mapping of a vast space of possible compositions. Theoretical predictions of CCAs properties are therefore of utmost importance.

In refractory CCAs, there is an inclination towards preparing the single-phase solid solution alloy, assuming that such alloy will be ductile, while intermetallic phases could cause embrittlement.[1] CALPHAD (CALculation of PHAse Diagrams) method is commonly used as a first estimation of phase composition in conventional alloys. Although CALPHAD databases are available even for CCAs, its usage is limited by the lack of reliable thermodynamic data and wide compositional space which need to be assessed. As a CALPHAD standard, only ternary phase diagrams are assessed during the database creation, which can decrease the accuracy of the predictions of alloys with higher number of elements.

Considering the main requirement—single phase composition—numerous simple semiempirical parameters were designed for CCAs for this purpose as proxies. The idea is that the prediction does not require a complete thermodynamic dataset and can be made with easy-to-obtain material constants, *e.g.* individual

JIŘÍ KOZLÍK, MARIANO CASAS LUNA, KRISTIÁN ŠALATA, JOSEF STRÁSKÝ, JOZEF VESELÝ, and ELIŠKA JAČA are with the Department of Physics of Materials, Faculty of Mathematics and Physics, Charles University, Ke Karlovu 5, 121 16 Prague, Czechia. Contact e-mail: jiri.kozlik@matfyz.cuni.cz FRANTIŠEK LUKÁČ and TOMÁŠ CHRÁSKA are with the Institute of Plasma Physics of the Czech Academy of Sciences, Za Slovankou 1782/3, 182 00 Prague, Czechia.

melting temperatures, atomic radii or ideal configuration entropy. We now briefly introduce and discuss several such parameters that aim at predictions of the presence of solid solution in the alloy.

### A. *Predictions of the Phase Composition*

The most basic thermodynamic parameters involve comparing the mismatch in atomic radius ($\delta r$) or mismatch in Allen's electronegativity ($\Delta\chi$). These parameters are easily computed using the rule of mixtures[2–4] using the same formula:

$$\delta r = \sqrt{\sum_{i=1}^{n} c_i\left(1-\frac{r_i}{\bar{r}}\right)^2} \quad [1]$$

$$\Delta\chi = \sqrt{\sum_{i=1}^{n} c_i\left(1-\frac{\chi_i}{\bar{\chi}}\right)^2} \quad [2]$$

where $c_i$ is the concentration of $i$th element, $r_i$ or $\chi_i$ are the given quantities (atomic radius or electronegativity) for given element, $\bar{r}$ or $\bar{\chi}$ are their weighted averages over all elements, $n$ is the number of elements present.

The region where a solid solution is said to form without any intermetallic compound is constrained by an atomic size mismatch ($\delta r$) of less than 6 pct and the electronegativity difference ($\Delta\chi$) ranging from 3 to 6 pct.[5] Similarly, Yurchenko *et al.* analyzed the formation of Laves phases and found that they form when $\Delta\chi$ exceeds 7 pct and $\delta r$ is greater than 5 pct.[6]

In 2012, Yang and Zhang developed one of the most widely utilized parameters for predicting solid solution formation in HEAs, known as the $\Omega$ parameter[7]:

$$\Omega\left(\overline{T_\mathrm{m}}\right) = \frac{\overline{T_\mathrm{m}}\Delta S_\mathrm{mix}}{|\Delta H_\mathrm{mix}|} \quad [3]$$

This parameter assesses the internal competition between the driving forces for solid solution formation (*i.e.,* the configurational entropy of the ideal solid solution, $\Delta S_\mathrm{mix} = -R\sum c_i \ln c_i$), and the forces driving the formation of intermetallic compounds $\Delta H_\mathrm{mix} = \sum_{i\neq j} 4\Delta H_{ij}c_i c_j$ (negative $\Delta H_\mathrm{mix}$) or the segregation of elements (positive $\Delta H_\mathrm{mix}$).[7] The melting point $\overline{T_\mathrm{m}}$ is calculated using a rule of mixture of individual melting points. According to Reference 5, all alloys exhibiting a single-phase microstructure can be found within the region where $\Omega \geq 1.1$ and $\delta r \leq 6.6$ pct. However, it has been demonstrated that these conditions are not the sole determinants for producing single-phase alloys, as the enthalpy of mixing $\Delta H_\mathrm{mix}$ is only a very crude approximation of the actual enthalpy of various phases which may be formed, *e.g.*, due to different crystal structure. Therefore, in many systems that include elements with a high chemical affinity for each other, the formation of intermetallic compounds or element segregation can occur, deviating from the predictions of these semiempirical methods.[8,9] Moreover, the $\Omega$ is calculated using the temperature of melting (or its approximate), which may be relevant for predicting phase composition of cast samples, but not for the samples which are annealed at another temperature. In this paper, we have calculated $\Omega$ using the actual annealing temperature $T_\mathrm{A}$ instead of $\overline{T_\mathrm{m}}$ as well.

Another parameter, $\phi$ parameter, introduced by Reference 10 aim on predicting the single-phase formation in HEAs:

$$\phi = \frac{\Delta S_\mathrm{mix} - \Delta S_\mathrm{H}}{|S_\mathrm{E}|} \quad [4]$$

where $\Delta S_\mathrm{H} = |\Delta H_\mathrm{mix}|/\overline{T_\mathrm{m}}$ and $S_\mathrm{E}$ is excess configurational entropy due to atomic size misfits as defined in Reference 11. Single-phase solid solutions should be stable for $\phi \geq 20$ pct. This threshold ensures that the entropic term at the melting point is greater than the enthalpy of mixing ($\Delta H_\mathrm{mix}$), and moreover the difference between these terms exceeds the excess configurational entropy ($S_\mathrm{E}$). Similarly to the previous parameter, the calculation can be and was performed using both $\overline{T_\mathrm{m}}$ and $T_\mathrm{A}$.

The last two parameters ($\Omega$ and $\phi$) predict the stability of a solid solution in CCAs by comparing the competing forces to stabilize the solid solution and the formation of intermetallics, *i.e.,* $\Delta S_\mathrm{mix}$ and $\Delta H_\mathrm{mix}$ terms, respectively. Nevertheless, this approach fails to consider that, regardless of how high $\Delta S_\mathrm{mix}$ or how low $\Delta H_\mathrm{mix}$ is, it cannot autonomously stabilize a solid solution if the Gibbs free energy of formation for a competing intermetallic compound ($\Delta G_\mathrm{IM}$) has a lower value.[12,13] For this reason, in 2016, King *et al.* introduced a new parameter, $\Phi$, to predict the formation of multi-principal element solid solutions.

$$\Phi(T) = \frac{\Delta G_\mathrm{mix}(T)}{-\left|\Delta G_\mathrm{IM}^{max}\right|} \quad [5]$$

They conducted a comparison between the Gibbs free energy for the formation of a fully disordered solid solution ($\Delta G_\mathrm{mix} = \Delta H_\mathrm{mix} - T\Delta S_\mathrm{mix}$) and the Gibbs free energy associated with the most probable binary intermetallic or element segregation $\Delta G_\mathrm{IM}^{max}$).[12]

Additionally, Senkov and Miracle[14] introduced another thermodynamic criterion—the complete enthalpy and entropy terms for the solid-solution and the formation of intermetallics. In this way, they considered that the $\Delta H_\mathrm{mix}$ and $\Delta S_\mathrm{IM}$ could have a significant impact on the stability of the solid solution. Their method assumed a linear relationship $\Delta H_\mathrm{IM} = \kappa_1\Delta H_\mathrm{mix}$, and also $\Delta S_\mathrm{IM} = \kappa_2\Delta S_\mathrm{mix}$. Subsequently, through a direct comparison of the Gibbs free energies of the competing phases and requiring $\Delta G_\mathrm{mix} < \Delta G_\mathrm{IM}$, a critical value of $\kappa_1 = \kappa_1^{cr}(T)$ can be found, determining the stability of the solid solution over the formation of the intermetallic compound at the calculated temperature.

$$\kappa_1 = \frac{\Delta H_\mathrm{IM}}{\Delta H_\mathrm{mix}} < -\frac{T\Delta S_\mathrm{mix}}{\Delta H_\mathrm{mix}}(1-\kappa_2) + 1 = \kappa_1^{cr}(T) \quad [6]$$

Table I. Overview the Semiempirical Parameters and Various Inputs Which They Use

| Parameter | Inputs | | | | | | | | | | References |
|---|---|---|---|---|---|---|---|---|---|---|---|
| | $r_i$ | $\chi_i$ | $\overline{T_m}$ | $T$ | $\Delta S_{mix}$ | $\Delta H_{mix}$ | $S_E$ | $\Delta G_{IM}^{max}$ | $\Delta H_{IM}$ | $\Delta S_{IM}$ | |
| $\delta r$ | x | | | | | | | | | | |
| $\delta\chi$ | | x | | | | | | | | | |
| $\Omega$ | | | x | opt. | x | x | | | | | 7 |
| $\phi$ | | | x | opt. | x | x | x | | | | 10 |
| $\Phi$ | | | x | x | x | x | | x | | | 12 |
| $\kappa_1$ | | | | x | x | x | | | x | x | 14 |

*Opt.* stands for *optional*, meaning that although the actual temperature is not included in the original definition, the parameter can be easily modified.

In their calculations, they presumed a partially ordered intermetallic phase, utilizing a sublattice arrangement of $(A,B)_1(C,D,E)_3$, where the configurational entropy represents more than 60 pct of the $\Delta S_{mix}$ in a five-component equiatomic ideal solid solution ($\kappa_2 = 0.6$).[14]

Finally, the discussed parameters are summarized in Table I, with the inputs they are using. The table should serve as a guide on parameter dependence, as well as a quick assessment of their complexity and number of inputs needed.

Given the increasing complexity of the semiempirical criteria and the necessity of obtaining many thermodynamic parameters anyway, the full thermodynamic calculation by CALPHAD (CALculation of PHAse Diagrams) methods can become actually competitive to the semiempirical parameters. Therefore, TCHEA4 database by ThermoCalc was also used in this paper for predictions of the phase composition. The aim of the paper is to compare the accuracy of both semiempirical and CALPHAD-based predictions of phase composition to the achieved experimental data for three RCCAs, and to compare mutually the efficiency of semiempirical parameters and CALPHAD predictions.

### B. *Preparation of RCCAs*

We will focus on three different equimolar RCCAs in this paper with the aim to achieve three different phase constitutions. All three RCCAs contain Al, Ti and Nb, while they differ in remaining two elements as highlighted below in bold:

(1) AlTiNb**MoV**—benchmark alloy, single phase microstructure expected[15,16]

(2) AlTiNb**TaZr**—alloy exhibiting intermetallic precipitates, resembling the microstructure of Ni superalloys[17,18]

(3) AlTiNb**MoCr**—low-density alloy, possible presence of Laves phase due to Cr causing embrittlement[15,19]

These alloys are typically prepared by casting. However, we propose a different approach, namely blended elemental powder metallurgy (BEPM). Its main benefit is a great variability of available chemical compositions, including preparation of compositionally graded materials.[20,21] The chemical homogeneity is achieved by a high-temperature sintering followed by homogenization treatment. In Reference 22, a parametric study was performed on Ti-35Nb-7Zr-5Ta (wt pct) alloy, showing that Ta, the most refractory element present, was fully homogenized after 1 hour at temperature 1500 °C to 1700 °C providing a hint for homogenization temperature and time.

## II. EXPERIMENTAL METHODS

### A. *Material Preparation*

Three equimolar mixtures were prepared from elemental powders and Al-Ti master alloy: AlTiNb**TaZr**, AlTiNb**MoV** and AlTiNb**MoCr**. Table II summarizes basic properties of the powders used in the experiments. The mixing of elemental powders was done using an in-house made rotational mixing device (a half-filled vial with the powder eccentrically placed within a rotating drum, 1 hour duration) immediately before filling the sintering die. No ball milling was used. We have previously shown that such mixing is capable of producing chemically homogenized alloys after subsequent thermal treatment described below.[23,24]

The field-assisted sintering (FAST), also known as spark plasma sintering (SPS), was performed in the SPS 10–4 furnace (Thermal Technology LLC) in a vacuum of the order of 1 Pa. The process was performed in two steps: heating up to 600 °C and a 20 min dwell to dissolve the Al, heating at the rate of 100 K/min up to the sintering temperature of 1300 °C, 1350 °C and 1400 °C for AlTiNb**TaZr**, AlTiNb**MoCr** and AlTiNb**MoV**, respectively. The temperatures were selected to be safely below any liquidus temperature in each system. The isothermal sintering was performed for 30 min under the pressure of 30 MPa. The samples were then cooled to the RT at the rate of 100 to 400 K/min. This condition is further referred to as the *as-sintered* condition.

Since this sintering program is insufficient for achieving chemical homogeneity with the elemental powders as a feedstock, the alloy bulks were subsequently *homogenized* at 1400 °C for 168 hour, followed by a furnace cool at a rate of 200 K/h maximum (lower at low temperatures). The homogenization was carried out under pure Ar atmosphere (99.9999 pct) and the samples were additionally wrapped in a Zr foil, acting as an oxygen getter. This condition will be referred to as the *homogenized* condition.

**Table II. The Results of a Chemical Analysis by a Carrier-Gas Hot Extraction (CGHE) Method in the *as-Sintered* Condition**

| Powder | Manufacturer | Size and Morphology | Oxygen (Wt Pct) | Nitrogen (Wt Pct) |
|---|---|---|---|---|
| Al50-Ti50 | American Elements | irregular (< 45 $\mu$m) | 0.931(3) | 0.219(6) |
| Nb | Alfa Aesar | irregular (< 45 $\mu$m) | 0.412(1) | 0.073(1) |
| Zr | TLS Technik | spherical (20–80 $\mu$m) | 0.175(1) | 0.009(1) |
| Ta | CAMEX | spherical (15–45 $\mu$m) | 0.076(8) | 0.007(1) |
| Mo | CAMEX | spherical (15–45 $\mu$m) | 0.037(12) | < 0.001 |
| Cr | Alfa Aesar | irregular (< 45 $\mu$m) | 0.636(3) | 0.004(1) |
| V | Goodfellow | irregular (< 45 $\mu$m) | 0.765(7) | 0.046(1) |

## B. *Experimental Characterization*

The oxygen and nitrogen contents in all samples were measured by a carrier gas hot extraction (CGHE) elemental analyzer (Bruker G8 Galileo) using 3 specimens of ~ 100 mg each.

The microstructural observation was done by scanning electron microscopy (SEM) using—FEI Quanta 200F and Zeiss Auriga Compact Crossbeam microscopes, both equipped with energy-dispersive X-ray spectroscopy (EDS) detectors (EDAX Octane), which were used for the measurements of chemical composition and element distribution. The lamella for TEM was prepared by Zeiss Auriga Compact Crossbeam and investigated using JEOL 2200FS transmission electron microscope.

Phase composition of alloys was determined by powder X-ray diffraction (PXRD) methods. The measurements were carried out on vertical powder $\theta$–$\theta$ diffractometer D8 Discover (Bruker AXS, Germany) using Cu K$\alpha$ radiation. Phase identification was done using X'Pert HighScore program with PDF-5 + database of crystalline phases and the quantitative Rietveld refinement was performed in TOPAS V5 (Bruker AXS, Germany).

The synergy of EDS and XRD, employed in this paper, is crucial in the research of CCAs. The symmetry of the intermetallic phases measured by XRD often corresponds to published structures, but the chemical composition and the site occupation do not. This affects the lattice parameters and the peak intensities. Therefore, a simple database lookup is not sufficient, as some parameters need to be relaxed to identify the phases correctly. These typically include site occupation and lattice parameters.

# III. RESULTS

At first, the results of the chemical analysis by CGHE are presented (Table III).

The oxygen content differs among the samples, which is caused by the purity of the original powders (as V and Cr powders contain significantly more oxygen than Zr, cf. Table II). During the homogenization, the oxygen content increased statistically significantly only in the AlTiNb**TaZr** sample. The N concentration is relatively low in all samples, the apparent decrease of N content in the AlTiNb**TaZr** sample is probably caused by its inhomogeneity in the *as-sintered* condition.

## A. *AlTiNbMoV Alloy*

The *as-sintered* microstructure of the AlTiNb**MoV** sample is shown in Figure 1, exhibiting a strong chemical heterogeneity, as expected. Undissolved Mo and Nb particles are clearly visible in the EDS maps, as well as clusters of TiAl particles.

The microstructure and chemical composition of the *homogenized* AlTiNb**MoV** alloys is presented in Figure 2. Several features can be identified: particles of a phase rich in Al and O (presumably $Al_2O_3$ from polishing and from the impurities in Al-Ti powder), Ti rich phase containing V as an alloying element and the chemically homogeneous matrix. Some pores are present as well as a result of dissolution and homogenization of elemental powder particles (the alloy having a different molar volume than the powder mixture).

The long-range chemical homogeneity was checked by an EDS linescan over the distance of 4.6 mm (presented in Figure 3). Good chemical homogeneity of the alloy was achieved by the long-term homogenization annealing, including homogenization of the slow-diffusing elements, such as Mo and Nb. The local deviations of the chemical composition correspond to the $Al_2O_3$ and Ti-rich particles described in the paragraph above.

To assess the chemical and phase composition properly, a combination of a quantitative EDS and XRD was used. The EDS results are presented in Table IV, along with the corresponding phase from the XRD. The XRD diffractogram of the *homogenized* AlTiNb**MoV** alloy (Figure 4) confirms that the major phase is a BCC phase. In order to quantify minor phases correctly, composition of each phase was taken from the EDS analysis (Table IV) and incorporated

Table III. The Results of a Chemical Analysis by a Carrier-Gas Hot Extraction (CGHE) Method in the *As-sintered* and *Homogenized* Conditions

| Material | Oxygen (Wt Pct) | | Nitrogen (Wt Pct) | |
|---|---|---|---|---|
| | As-sintered | Homogenized | As-sintered | Homogenized |
| AlTiNb**MoV** | 0.332(23) | 0.328(33) | 0.044(1) | 0.065(10) |
| AlTiNb**TaZr** | 0.192(17) | 0.226(7) | 0.059(1) | 0.048(5) |
| AlTiNb**MoCr** | 0.323(12) | 0.343(11) | 0.068(2) | 0.062(4) |

The numbers in parentheses are standard deviations at the last digit, three specimens were measured for each sample (except for *homogenized* AlTiNb**TaZr**, where only two specimens were measured).

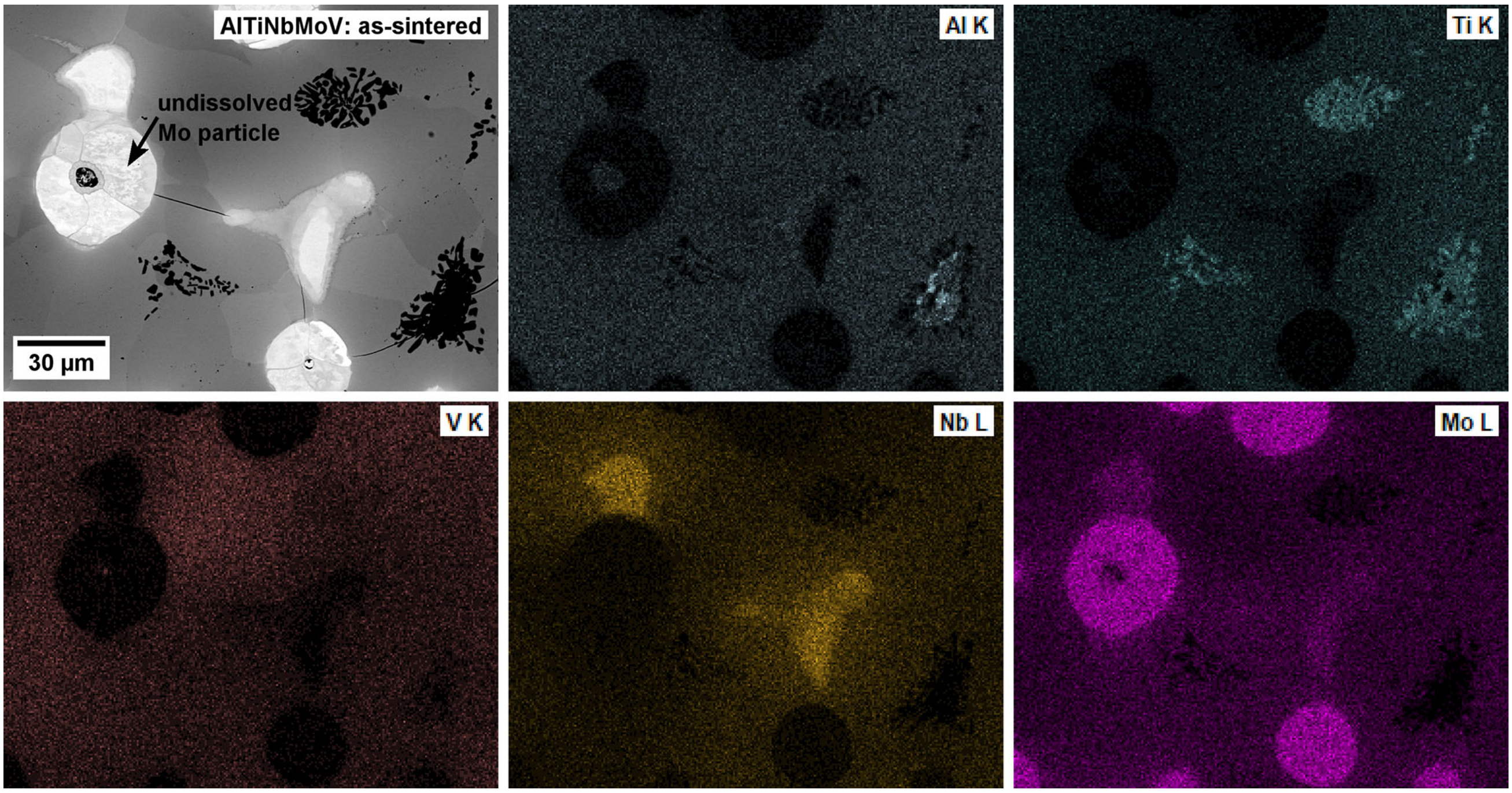

Fig. 1—Microstructure and chemical composition of the *as-sintered* AlTiNb**MoV** alloy.

into Rietveld refinement procedure with accurate site occupancies.

The lattice parameter of the AlTiNb**MoV** matrix is 3.1726 Å and a minor phase rich in Ti is of the FCC structure, having the lattice parameter of 4.2477 Å. The FCC phase can be associated with the Ti-rich particles visible in the EDS map in Figure 2, since our detailed EDS analysis revealed high concentration of O, N and C, corresponding to titanium (carb)oxynitrides having FCC structure and with a variable ratio of the light elements. Titanium rich nitrides are also natural for heat treated RCCA alloys with gas impurities, as pointed out in Reference 25.

## B. *AlTiNbTaZr Alloy*

The *as-sintered* microstructure of the AlTiNb**TaZr** sample is shown in Figure 5. Similarly to the AlTiNb**MoV**, the *as-sintered* sample is strongly heterogeneous with undissolved particles (mostly Zr, Nb and Ta).

The microstructure and chemical composition of the homogenized AlTiNb**TaZr** alloys is presented in Figure 6. Two-phase microstructure is observed: a matrix containing all elements and precipitates rich in Al and Zr (the higher Nb content in the precipitates is in fact a false signal coming from the Zr L$\beta$ and Nb L$\alpha$ peak overlaps; a more accurate quantification is presented further down in the text). Two classes of precipitates can be observed with a different morphology—equiaxed precipitates of up to 100 $\mu$m in diameter and precipitates at the grain boundaries of the matrix.

The elemental distribution was measured by EDS line scan, and the resulting compositional profile is presented in Figure 7. It shows that initial elemental powder particles were dissolved during thermal treatment as there are no chemical variations at the length-scale of these initial particles. Chemical composition of all elements is evenly distributed over the length of the line scan (> 4.5 mm), except for Al- and Zr-rich secondary phase precipitates shown already in Figure 6. It should be pointed out that even the initial particles of the refractory and slow-diffusing Ta were successfully dissolved in the *homogenized* alloy.

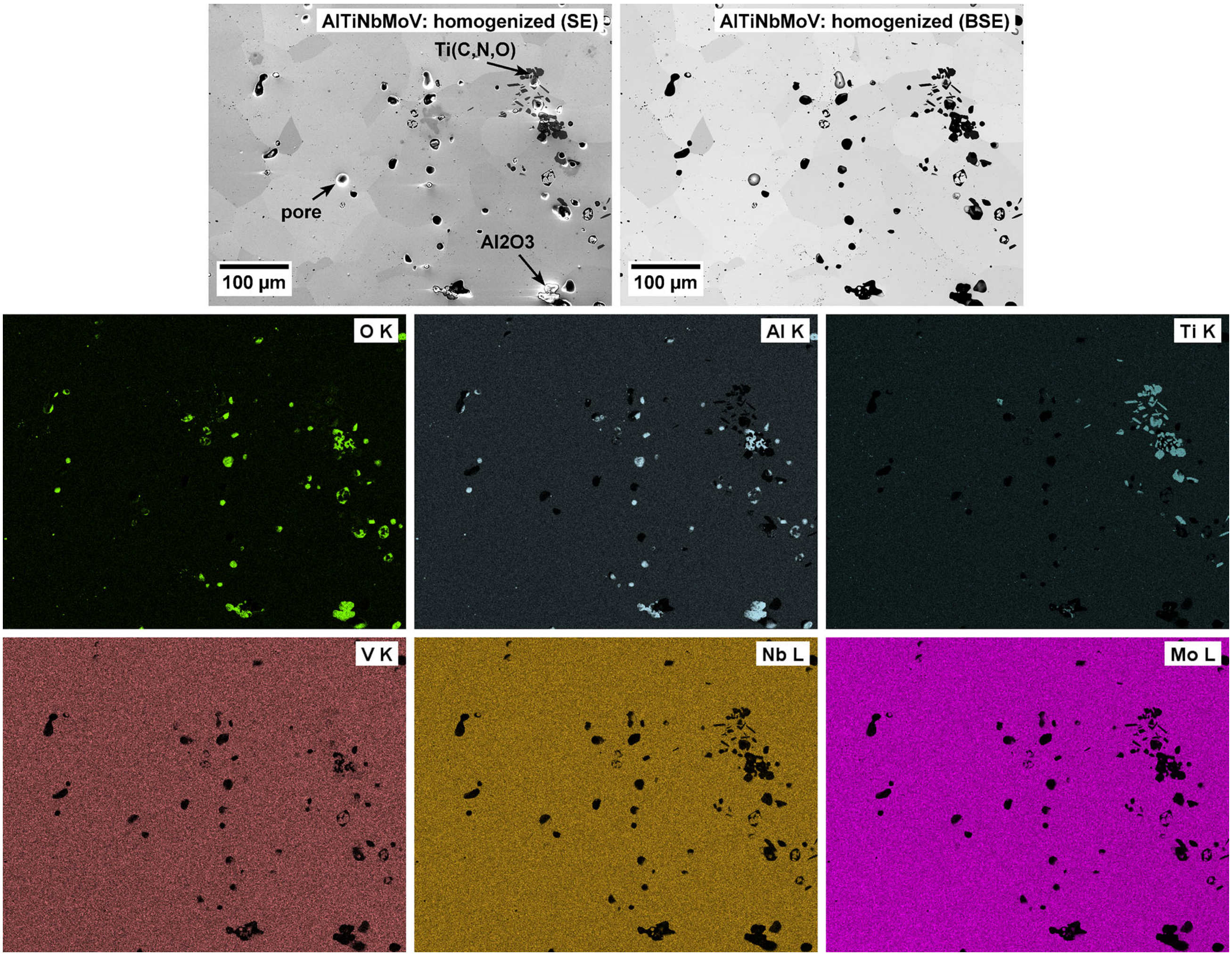

Fig. 2—Microstructure of the *homogenized* AlTiNb**Mo**V and corresponding EDS maps.

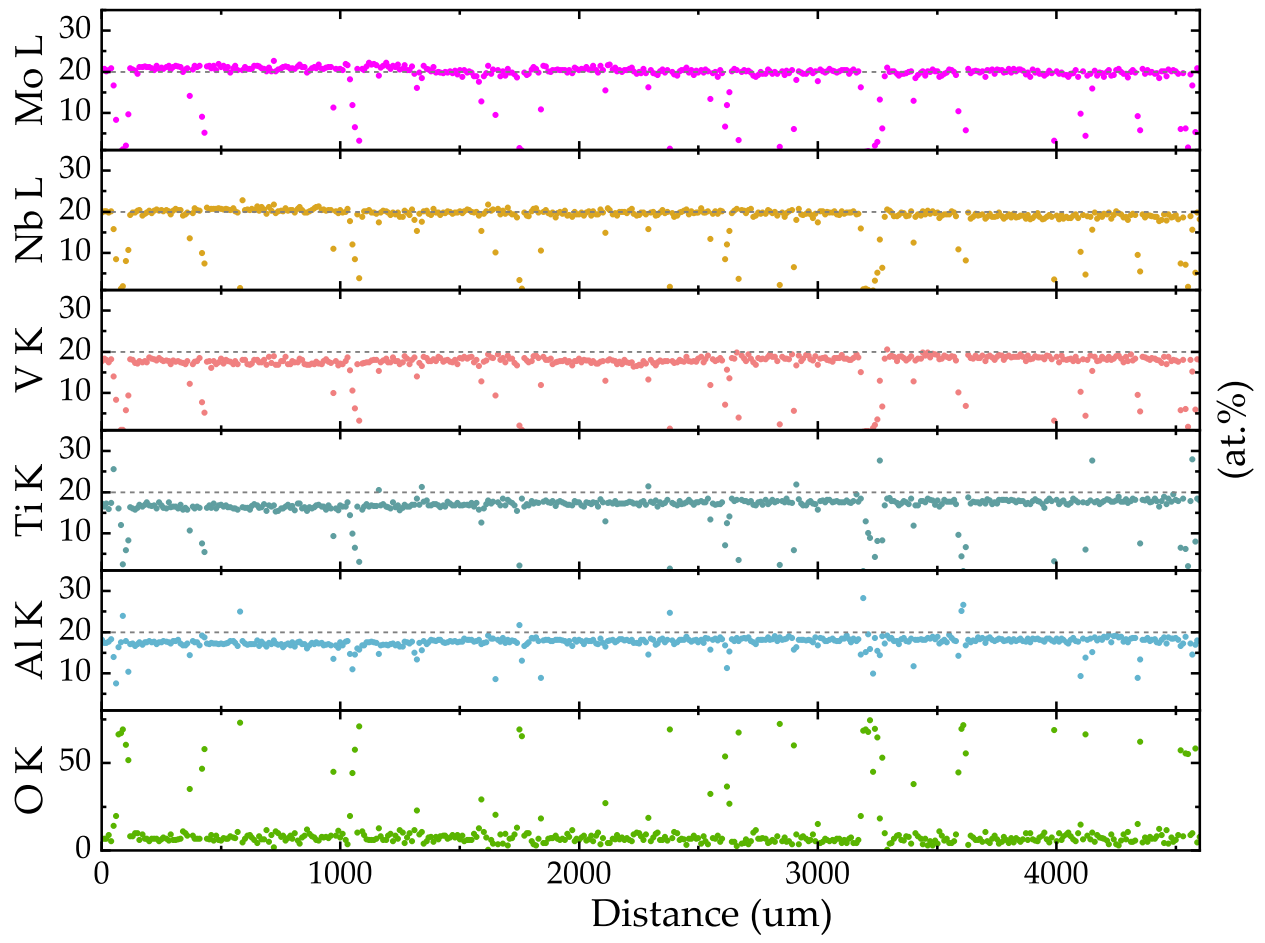

Fig. 3—EDS line scan of the *homogenized* AlTiNb**Mo**V alloy over the distance of 4.6 mm, showing chemical homogeneity at the scale of the former powder particles. The concentrations should be regarded as qualitative only, as the dwell time (and the number of counts) per point is too low for an accurate quantification. The dashed lines mark the nominal composition of the alloy. The outlying points correspond to the ceramic particles.

Large GB precipitates rich in Al and Zr (top right corner in Figure 8) and decomposition to two BCC phases (marked by arrows in Figure 8) can be clearly identified. Such microstructures were observed in several similar alloys containing Al and Zr annealed at 1200 °C for 24 hours and cooled at the rate of 10 K/min.[26] Very fine precipitates were found within the AlTiNb**TaZr** matrix, shown in Figure 8 (black arrow). Due to their very small size (50–200 nm), we assume that these precipitates formed during the final furnace cool after the homogenization and are not expected to be in a thermodynamic equilibrium at annealing temperature. The EDS analysis of the precipitates is also impossible because of the small size compared to the X-ray interaction volume in SEM.

Table IV. Chemical and Phase Composition of the *Homogenized* AlTiNbMoV Alloy as Determined by EDS and XRD

| | XRD | | EDS | | | | | | | |
|---|---|---|---|---|---|---|---|---|---|---|
| | Structure | Lattice Parameters (Å) | C (At. Pct) | N (At. Pct) | O (At. Pct) | Al (At. Pct) | Ti (At. Pct) | Nb (At. Pct) | V (At. Pct) | Mo (At. Pct) |
| Matrix | BCC | $a$ = 3.1726 | — | — | — | 18.1 | 19.2 | 20.5 | 18.9 | 23.3 |
| Ti(C,N,O) | FCC | $a$ = 4.2477 | 13.4 | 13.6 | 9.3 | 0.5 | 60.3 | 1.7 | 0.8 | 0.5 |
| $Al_2O_3$ | $R\bar{3}c$ | $a$ = 4.7609 $c$ = 13.007 | — | — | 60.1 | 33.5 | 2.0 | 1.6 | 1.2 | 1.6 |

The Ti(C,N,O) chemical composition should be regarded as qualitative only, as the quantification of light elements overlapping with Ti is not precise.

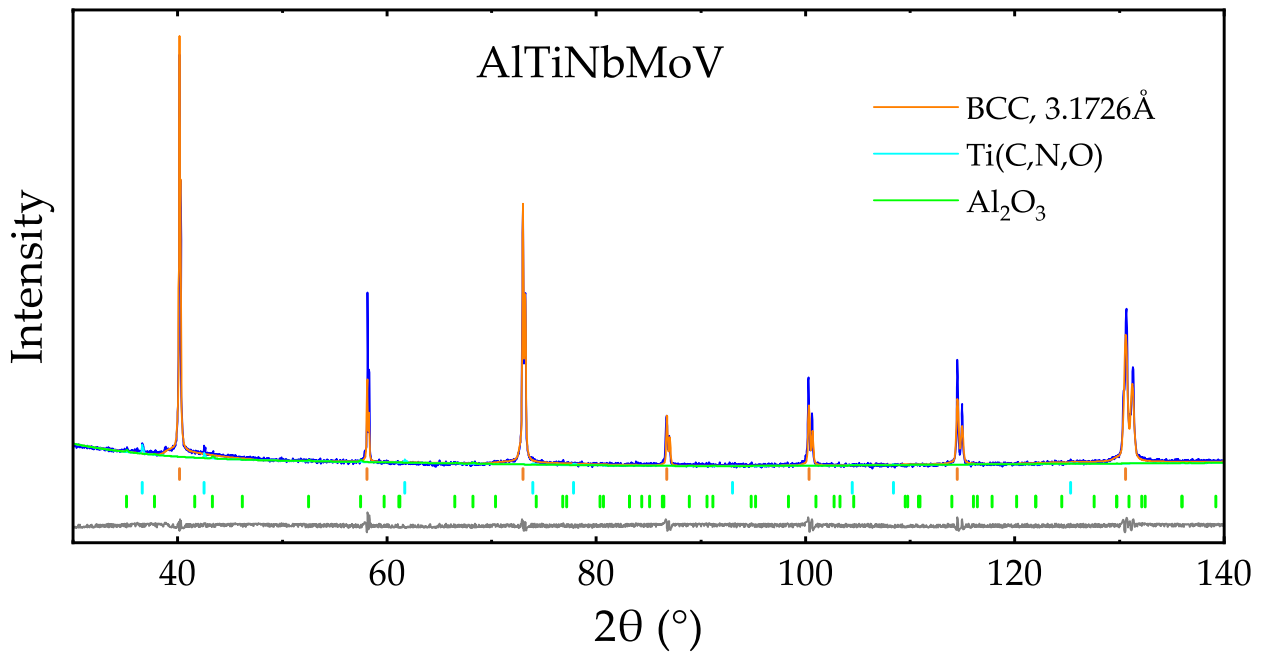

Fig. 4—X-ray diffractogram of the *homogenized* AlTiNb**MoV** alloy. Two phases were identified: a dominant solid solution BCC phase and a minor Ti(C,N,O) FCC phase.

Similarly to AlTiNb**MoV** sample, a combination of EDS data and XRD were used for a detailed phase and chemical analysis.

XRD data (Figure 9) reveal that the lattice parameter of major BCC phase in AlTiNb**TaZr** sample is 3.039 Å. The matrix is depleted of Al and Zr (compared to the nominal alloy composition) as Al- and Zr-rich secondary phase is formed. A minor phase with BCC structure, the lattice parameter of 3.348 Å and very broad peaks was identified as well. The third phase is a phase isomorphous with the $Mn_5Si_3$-type structure. It is expected to be an $Al_3Zr_5$-based phase (as denoted in Figure 8) with hexagonal structure of space group P63/mcm. However, the fitted lattice parameters are very far from the table values found in the PDF5 + database ($a$ = 8.184 Å, $c$ = 5.702 Å, PDF# 00-012-0674). Therefore, the EDS results from Table V were incorporated into the occupancy of the Zr sites by substitution of Zr atoms by proportional numbers of Ti, Nb and Ta atoms found by EDS. The Al sites are occupied by Al atoms as the measured percentage of Al of 33.9 at. pct from Table V is very close to the theoretical value of 37.5 at. pct. This is natural for the $Al_3M_5$ structures usually found as combination of a transition metal and an element from III.A and IV.A groups of the periodic table.

## C. *AlTiNbMoCr Alloy*

The *as-sintered* microstructure of the AlTiNb**MoCr** sample is shown in Figure 10. Similarly to the AlTiNb**MoV**, the *as-sintered* sample is strongly heterogeneous with undissolved Nb and Mo particles.

The *homogenized* alloy micrographs and EDS maps are shown in Figure 11. Similarly to AlTiNb**MoV**, there are Al- and O-rich areas (presumably $Al_2O_3$) and some Ti-rich particles found in the matrix. The EDS line scan confirms good homogenization of the sample even of this system, as shown in Figure 12. Local fluctuations of the chemical composition correspond to the $Al_2O_3$ particles.

The detailed SEM micrograph of a triple point is shown in Figure 13(a). Due to the small size of the precipitates, a TEM lamella was prepared from the grain boundary. The phases are shown in detail in Figure 13(e), along with the chemical composition of the area. The corresponding SAED patterns are then shown in Figures 13(b) through (d).

The SAED pattern of the matrix has revealed a partial B2 ordering, as demonstrated by additional diffraction spots at the positions forbidden for BCC structure. The grain boundary phase (dark in Figure 13(a)) corresponds to the hexagonal C14 Laves phase ($MgZn_2$ prototype) and it is significantly enriched with Cr and depleted of Ti and Mo. The second phase was identified as a cubic A15 phase ($Cr_3Si$ prototype) and its composition is close to that of the matrix (only slightly depleted of Ti).

The XRD diffractogram is shown in Figure 14 and its extract is summarized in Table VI along with the EDS data. Major phase in AlTiNb**MoCr** is a BCC phase with the lattice parameter of 3.149 Å. The additional peaks resulting from the ordering were not detected, likely due to their low intensity. Minor phases were identified as the corundum $Al_2O_3$ phase, an FCC phase with the lattice parameter of 4.2636 Å (presumably FCC Ti(C,N,O) phase as in AlTiNb**MoV** alloy) and a cubic phase A15 (space group $Pm\bar{3}n$) isomorphous with the $Cr_3Si$ phase with the lattice parameter of 5.0309 Å. According to EDS, one sublattice of this phase was set to be occupied by Al atoms and the other sublattice comprises of Cr, Nb, Ti, Mo atoms in the ratio determined by EDS in Table VI. The C14 Laves phase was not detected by XRD due to its low volume fraction.

A similar microstructure with the GB phases was observed in Reference 27 after annealing at 1300 °C for 20 hours and cooling at 4.2 K/min (close to that used in this work), but the authors were unable to identify them completely. The observed minor phases arise either from impurities ($Al_2O_3$ and Ti(C,N,O)) or their particles are small (C14 Laves and A15 $Cr_3Si$), suggesting that they were not formed during the long homogenization treatment (as the particles would grow significantly after 168 hours, cf. the intermetallic phases in AlTiNb**TaZr** in Figure 6 reaching tens of $\mu$m). Therefore, for the sake of comparison with the thermodynamic predictions, the AlTiNb**MoCr** alloy will be regarded as a single-phase material at 1400 °C.

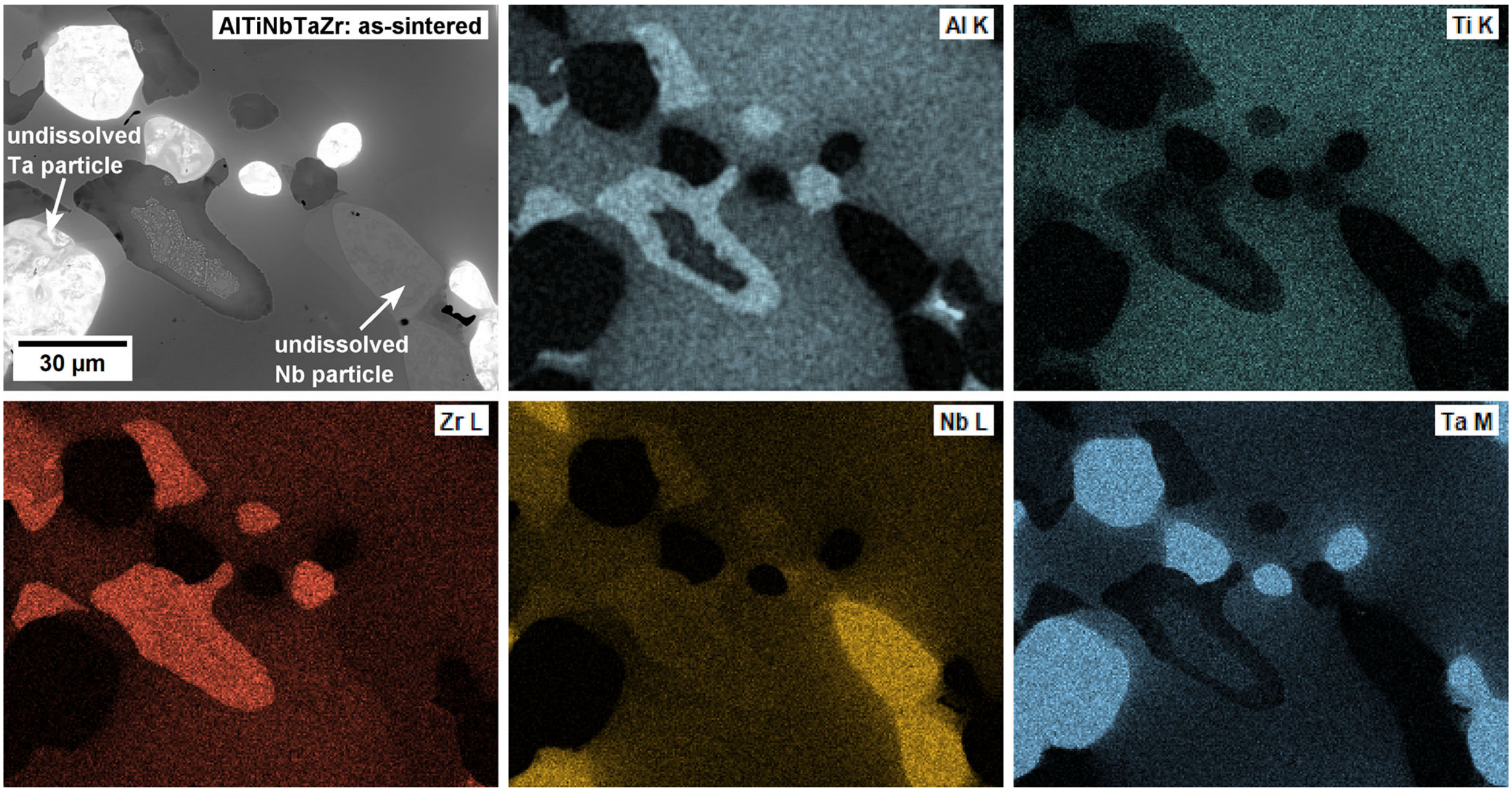

Fig. 5—Microstructure and chemical composition of the *as-sintered* AlTiNb**TaZr** alloy.

## IV. DISCUSSION

The non-homogeneity of the *as-sintered* alloys is not surprising, as the sintering at 1300–1400 °C for 30 min is not sufficient for a formation of RCCA from initial elemental powder particles. Based on our previous experience, sintering temperature of at least 1500 °C is necessary to achieve homogeneous chemical distribution of Ti–Nb–Zr alloy (*i.e.* alloy without Ta).[24] This fact led us to apply the homogenization treatment at 1400 °C for 168 hours, which proved to be successful, yielding a full dissolution of the elemental powders over entire sample and formation of RCCA while keeping the O and N contamination low. In addition, Ti and Al acted as strong N and O getters in AlTiNb**MoV** and AlTiNb**MoCr** alloys resulting in formation of oxides.

In the next part of the discussion, the accuracy of the phase predictions will be discussed, first the semiempirical parameters and consequently the CALPHAD calculation. The ceramic phases (Ti(C,N,O) and $Al_2O_3$) detected by EDS and XRD are disregarded for the sake of phase composition comparison, as they do not significantly influence the competition between the solid solution and the formation of intermetallic phases and, moreover, both semiempirical parameters and CALPHAD do not take them into account.

### A. *Semiempirical Parameters*

The semiempirical parameters which were used for the predictions of phase composition are summarized in Table VII. The parameters were calculated using the Eqs. [1] through [6]. The values of $\Phi$ were calculated using the original script kindly provided by the originators.[12] Some are temperature dependent—their values were calculated at the estimated melting temperature $\overline{T_m}$ (when included in the original parameter definition) and at the actual annealing temperature $T_A = 1673$K for comparison. The calculation of $\kappa_1^{cr}(T) - \kappa_1$ parameter requires a selection of a competing intermetallic phase. $(Al)_3(Nb,Ti,Ta,Zr)_5$ model was selected for the

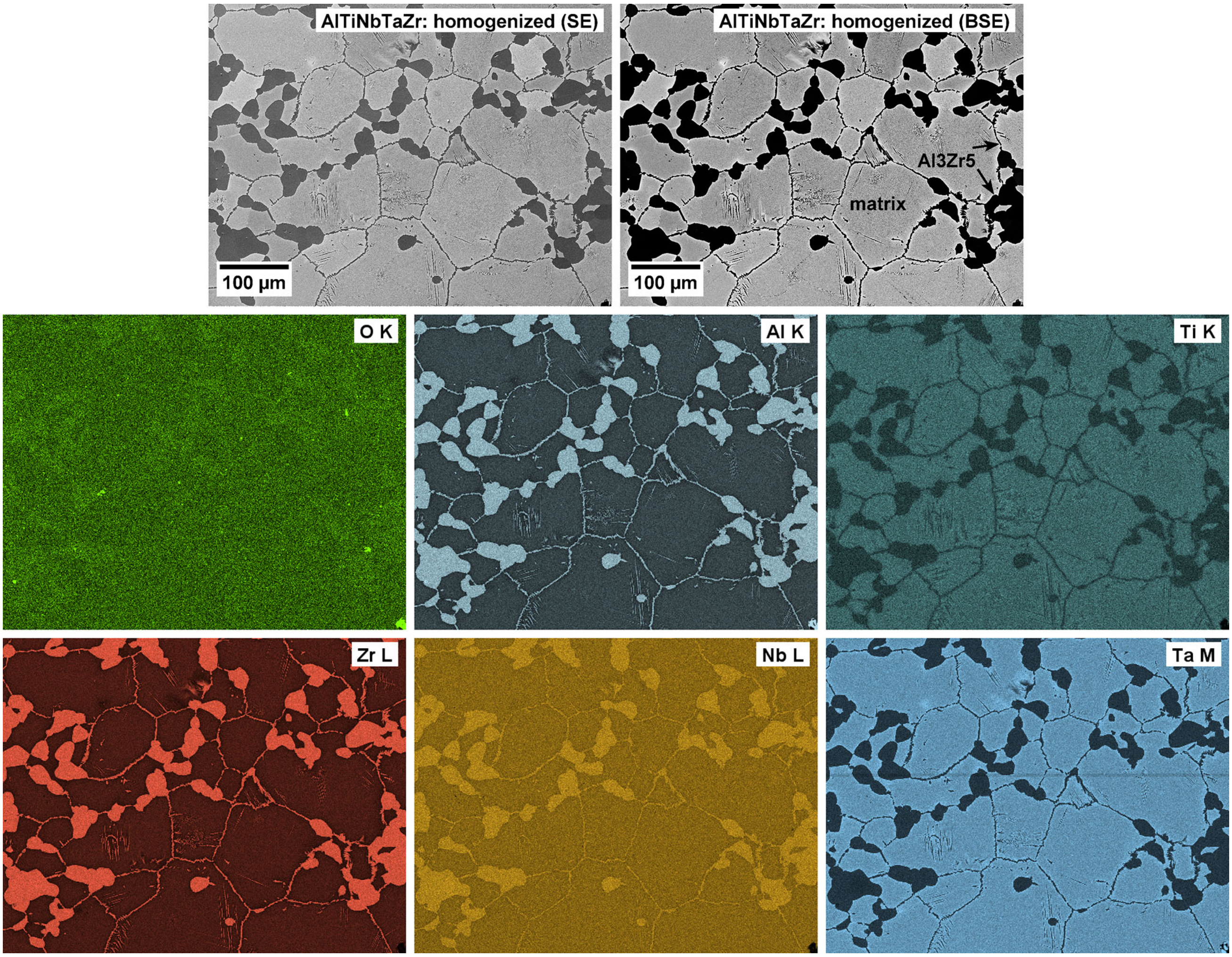

Fig. 6—Microstructure of the *homogenized* AlTiNb**TaZr** and corresponding EDS maps. The Nb map is not accurate, as the Nb L$\alpha$ line overlaps with the Zr L$\beta$ line and a false Nb signal is picked up in Zr-rich regions.

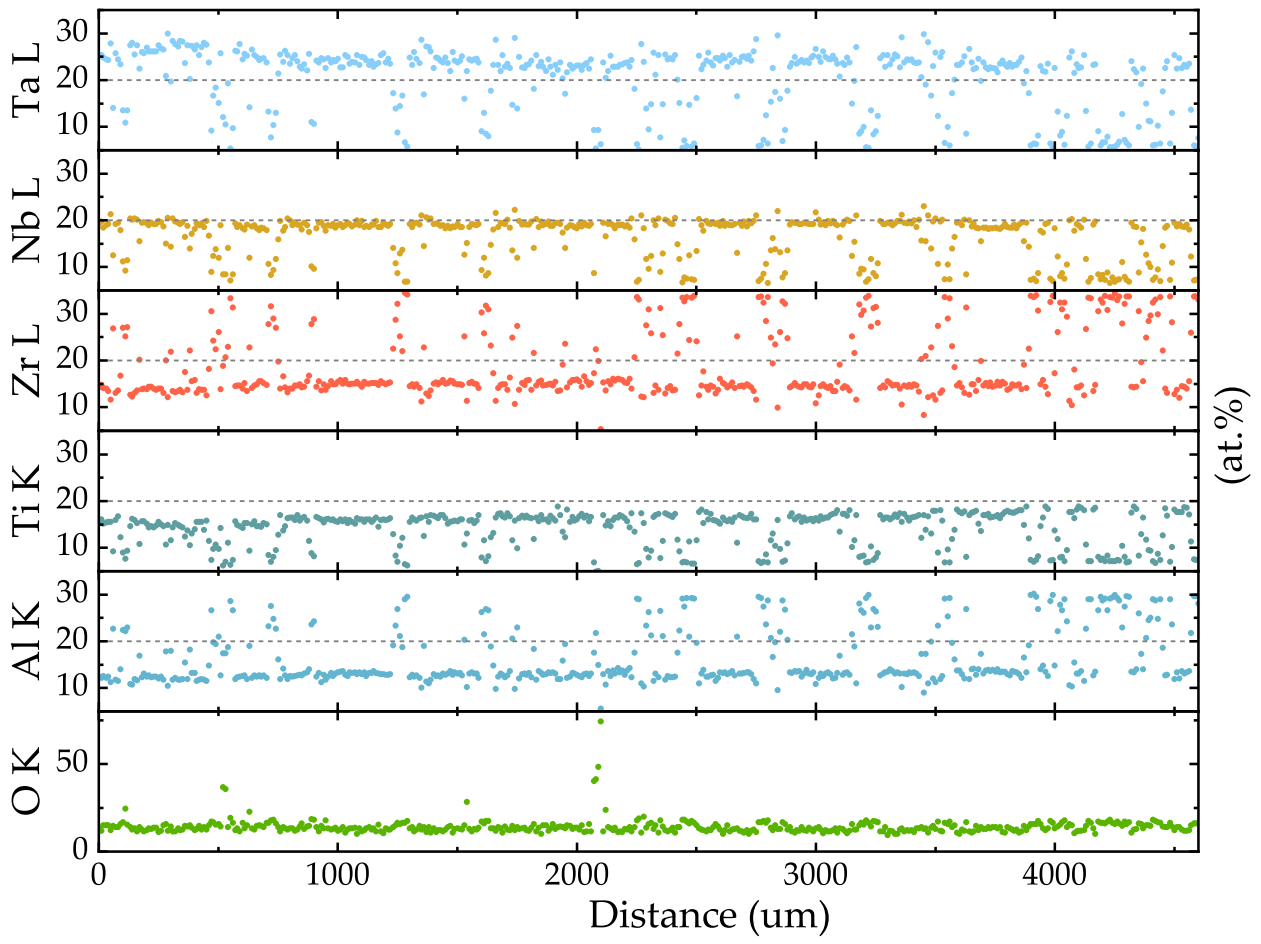

Fig. 7—EDS line scan of the AlTiNb**TaZr** alloy over the distance of 4.6 mm, showing chemical homogeneity at the scale of the former powder particles. The concentrations should be regarded as qualitative only, as the dwell time (and the number of counts) per point is too low for an accurate quantification. The outlying points correspond to the $Al_3Zr_5$-based intermetallics.

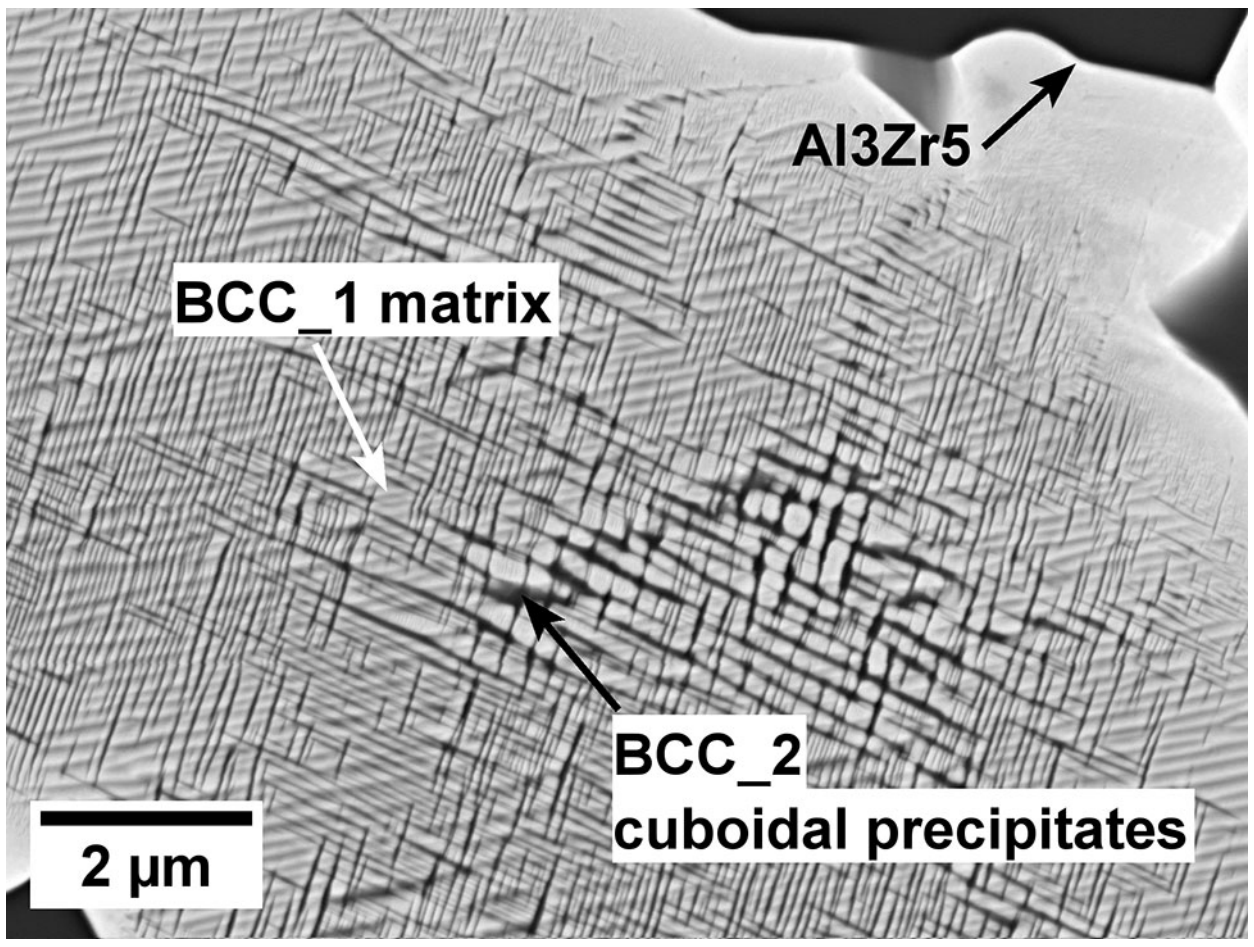

Fig. 8—SEM BSE micrograph of fine precipitates (dark) within the *homogenized* AlTiNb**TaZr** matrix (bright) arranged in a cuboidal pattern. $Al_3Zr_5$-based intermetallic phase is visible in the top right corner.

Table V. Chemical and Phase Composition of the AlTiNbTaZr Alloy as Determined by EDS and XRD

| | XRD | | EDS | | | | |
|---|---|---|---|---|---|---|---|
| | Structure | Lattice Parameters (Å) | Al (At. Pct) | Ti (At. Pct) | Nb (At. Pct) | Zr (At. Pct) | Ta (At. Pct) |
| Matrix | BCC_1 | $a = 3.039$ | 16.0 | 18.8 | 21.6 | 17.8 | 25.9 |
| | BCC_2 | $a = 3.348$ | | | | | |
| $Al_3(Zr,Ti,Nb,Ta)_5$ | P63/mcm | $a = 8.0710$ | 33.9 | 10.8 | 8.8 | 39.6 | 6.9 |
| | | $c = 5.4667$ | | | | | |

The two BCC phases could not be distinguished by EDS and the results represent an average composition over the area (cf. Fig. 8).

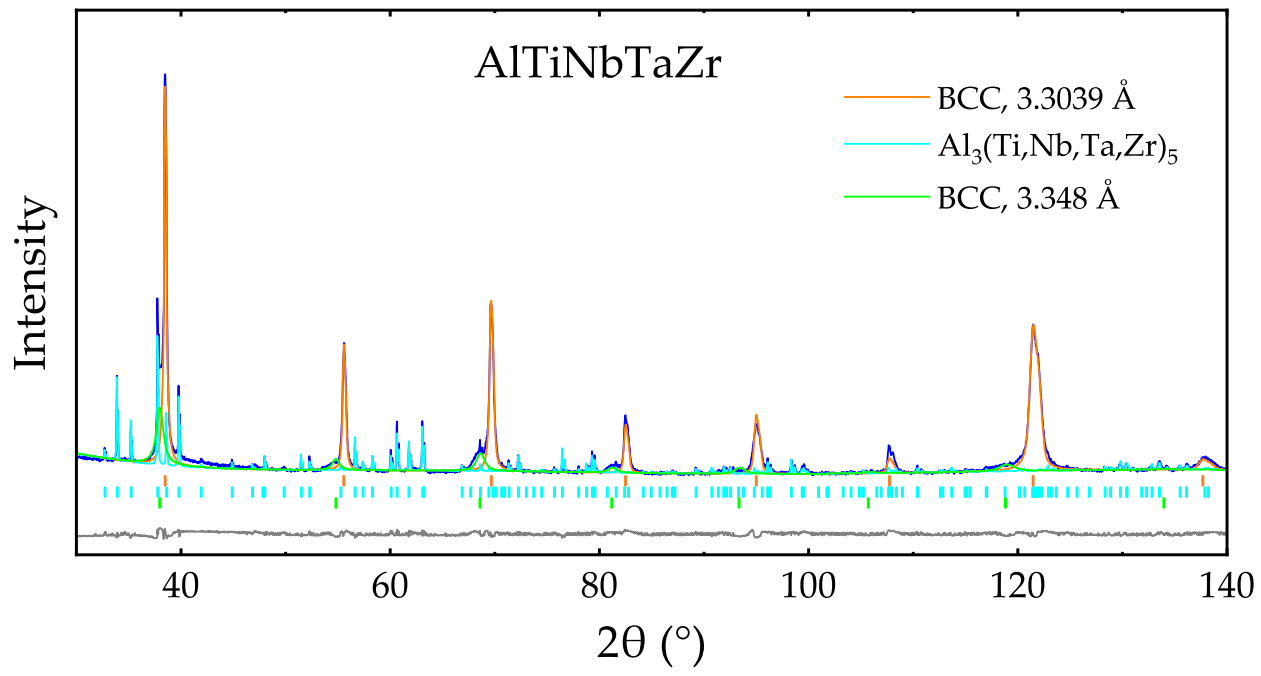

Fig. 9—X-ray diffractogram of the AlTiNb**TaZr** alloy. Three phases were identified: a dominant solid solution BCC phase, minor BCC phase with broad peaks and the $Al_3Zr_5$-based intermetallic.

AlTiNb**TaZr** alloy, resulting in $\kappa_2 = 0.53$. For the other two alloys, a default estimation of $\kappa_2 = 0.6$ was used due to the absence of any natural candidate for the competing intermetallic phase. Table VIII finally presents the comparison of the predictions with the experimental observations.

All studied systems met the criteria of having a low atomic size mismatch and higher system entropy compared to the enthalpy of mixing ($\Omega + \delta r$ criterion) at both temperatures, $\overline{T_m}$ and $T_A$ and solid solution was predicted in all of them. The formation of SS in the AlTiNb**TaZr** case was predicted incorrectly at 1400 °C, although it correctly showed the qualitative trend with Ω parameter being closer to the intermetallic range.

Using the temperature insensitive $\Delta H_{mix} + \delta r$ criterion, the presence of intermetallics was predicted in all systems, but they were found only in the AlTiNb**TaZr** alloy.

The simplest criterion, average electronegativity difference and atomic size mismatch ($\Delta\chi + \delta r$), which does not require any thermodynamic input, correctly predicted BCC structure of the AlTiNb**MoV** alloy and the presence of intermetallics in the AlTiNb**TaZr** alloy, but incorrectly predicted phase composition of the AlTiNb**MoCr** alloy.

The $\phi$ parameter correctly predicted the phase composition of the AlTiNb**MoV** and AlTiNb**TaZr** alloys when used in the original form with the $\overline{T_m}$. When the solid solution stability is decreased by using the actual annealing temperature of 1400 °C, intermetallic phases are predicted for the AlTiNb**MoV** alloy as well, while not observed. Nevertheless, qualitative trends were captured correctly, with the $\phi$ parameter being highest in the single-phase AlTiNb**MoV** alloy.

Using the Φ parameter, a microstructure with the presence of intermetallic phase was predicted for both temperatures used for calculation, which is correct only for the AlTiNb**TaZr** alloy. The enthalpy of formation for the binary Al-Ti or Al-Zr intermetallic compounds was sufficiently high to predict the formation of these intermetallic compounds rather than a single-phase solid solution. However, this parameter needs to be carefully considered in alloys with high aluminum content, as the short range ordering can help stabilizing the solid solution in the system, as reported for some exceptions to this model.[13]

Finally, the phase composition predicted by $\kappa$ parameters is incorrect for all three alloys, although it is close to the critical value in the case of AlTiNb**TaZr**. In the case of the prediction using $\kappa$ parameters, the selection of the competing intermetallic phase is important part of the calculation, which requires some previous experience with the system. The intermetallics stoichiometry influences the configurational entropy term $\Delta S_{IM}$ and consequently the calculated value of $\kappa_2$ (recall that $\kappa_2 = \Delta S_{IM}/\Delta S_{mix}$). For example, the stoichiometries $(A,B)_1(C,D,E)_1$ or $(A,B)_1(C,D,E)_3$ would lead to $\kappa_2$ of 0.59 and 0.66, respectively. However, this would not change the prediction results.

None of the presented semi-empirical parameters has correctly predicted the phase composition of all three prepared alloys. Some parameters (Ω and ϕ) were at least able to qualitatively capture the trends between solid solution and intermetallics formation. However, the practical applicability of these semiempirical criteria for achieving single phase solid solution is limited—the simple parameters are unable to capture the thermodynamics of the system fully as most of them have been developed on the basis of limited amount of experimental data, moreover usually of as-cast alloys. On the other hand, the more complex parameters presented require additional input and knowledge about the studied alloying system, such as constitution of the possible intermetallic phases. Compared to such requirements, running a full CALPHAD calculation for given composition and temperature does not add significantly more difficulty. Similar undecisive phase prediction by the semi-empirical parameters in RCCAs has been reported also for other studied CCAs, showing their deficiency in predictions with accuracy below 70 pct in most of the cases.[29,30]

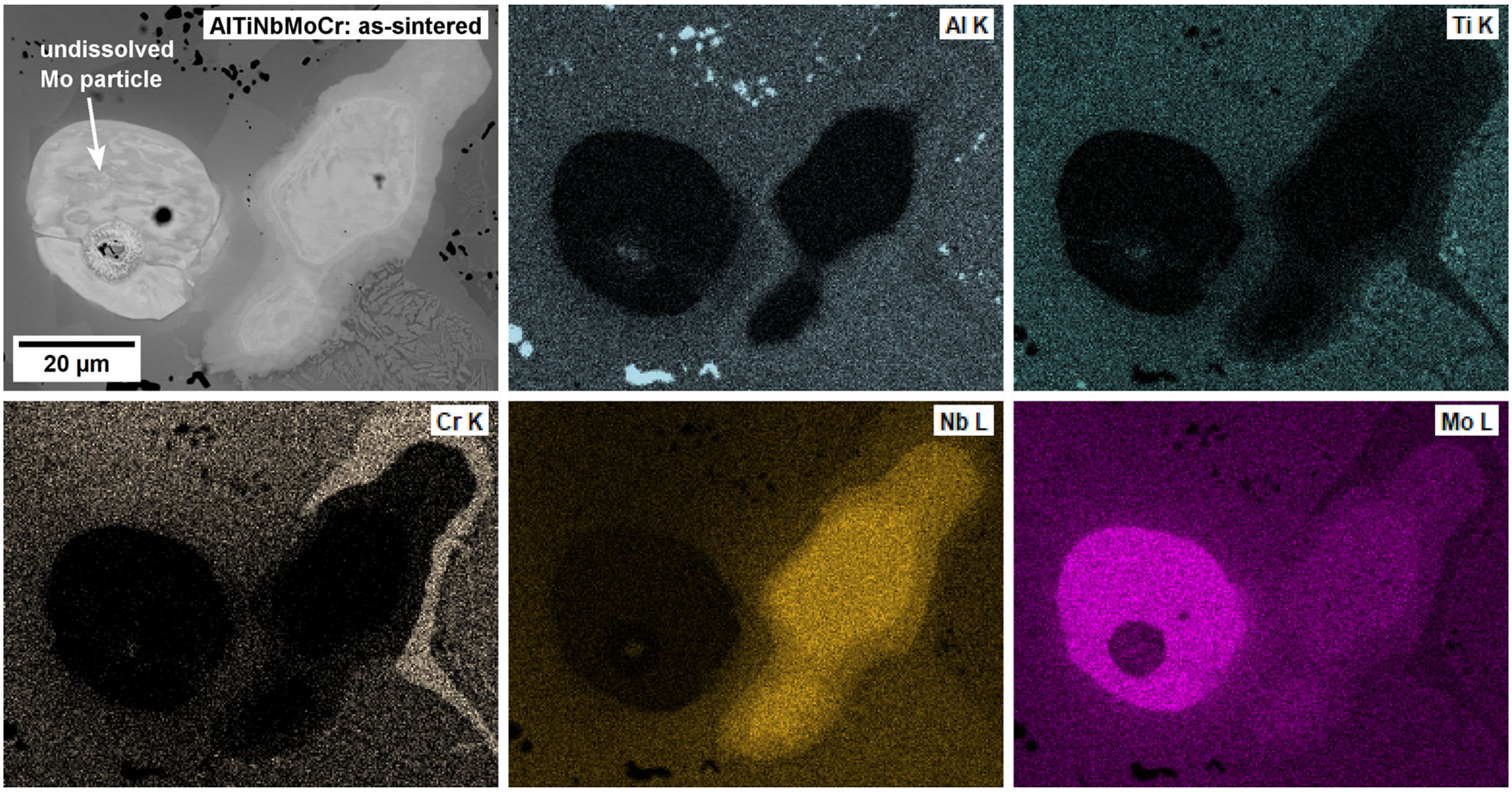

Fig. 10—Microstructure and chemical composition of the *as-sintered* AlTiNb**MoCr** alloy.

## B. *CALPHAD Calculations*

The CALPHAD calculations were performed using the ThermoCalc software with the TCHEA4 database. This database does not include thermodynamic information about the impurity elements, such as O and N, so those were excluded from the calculation. Equimolar nominal composition was assumed for all alloys. An equilibrium calculation was performed for the homogenization temperature of 1400 °C and compared with the phases measured by XRD and EDS. The comparison of the theoretical calculations and the experiment is presented in For each criterion, the condition for single-phase solid solution is presented below. The entries in bold are in accordance with the experimental observations. The actual annealing temperature $T_A$ = 1673K for all alloys Table IX. Only the predictions for the (inter)metallic phases were assessed.

The CALPHAD method correctly predicted the single-phase structure of the AlTiNb**MoV** alloy. In the case of AlTiNb**TaZr**, the calculation is mostly correct, as both matrix and $Al_3Zr_5$-based intermetallics are predicted. However, the definition of *AL3ZR5_D8M* phase in the TCHEA4 database is $(Al)_3(Zr,Ti)_5$, while the identified phase is more complex in terms of chemical composition as it contains significant concentrations of Nb and Ta. The presence of the fourth and fifth element in the calculated phase, in this case Ta and Nb, is not possible because the database is assessed using binary and ternary diagrams only. Since this structure was not found in any of the remaining ternary combinations of the system, it was assumed that only Al, Ti and Zr can be present.

For the AlTiNb**MoCr** alloy, CALPHAD predicts a two-phase microstructure consisting of BCC matrix and Cr-rich Laves phase. The Laves phase was detected only in the form of micrometric precipitates at grain boundaries in significantly smaller volume fraction than the predicted 20 pct (Figure 15). Considering that precipitates of this size have most probably formed during the cooling, it suggests that the driving force for Laves phase precipitation is too low at 1400 °C. This is consistent with Reference 27, where they observed Laves phase after annealing at 1100 °C and 1200 °C, but not at 1300 °C. It can be concluded that CALPHAD appropriately predicts that Laves phase may form in the studied alloy, but it overestimates its stability at high temperatures.

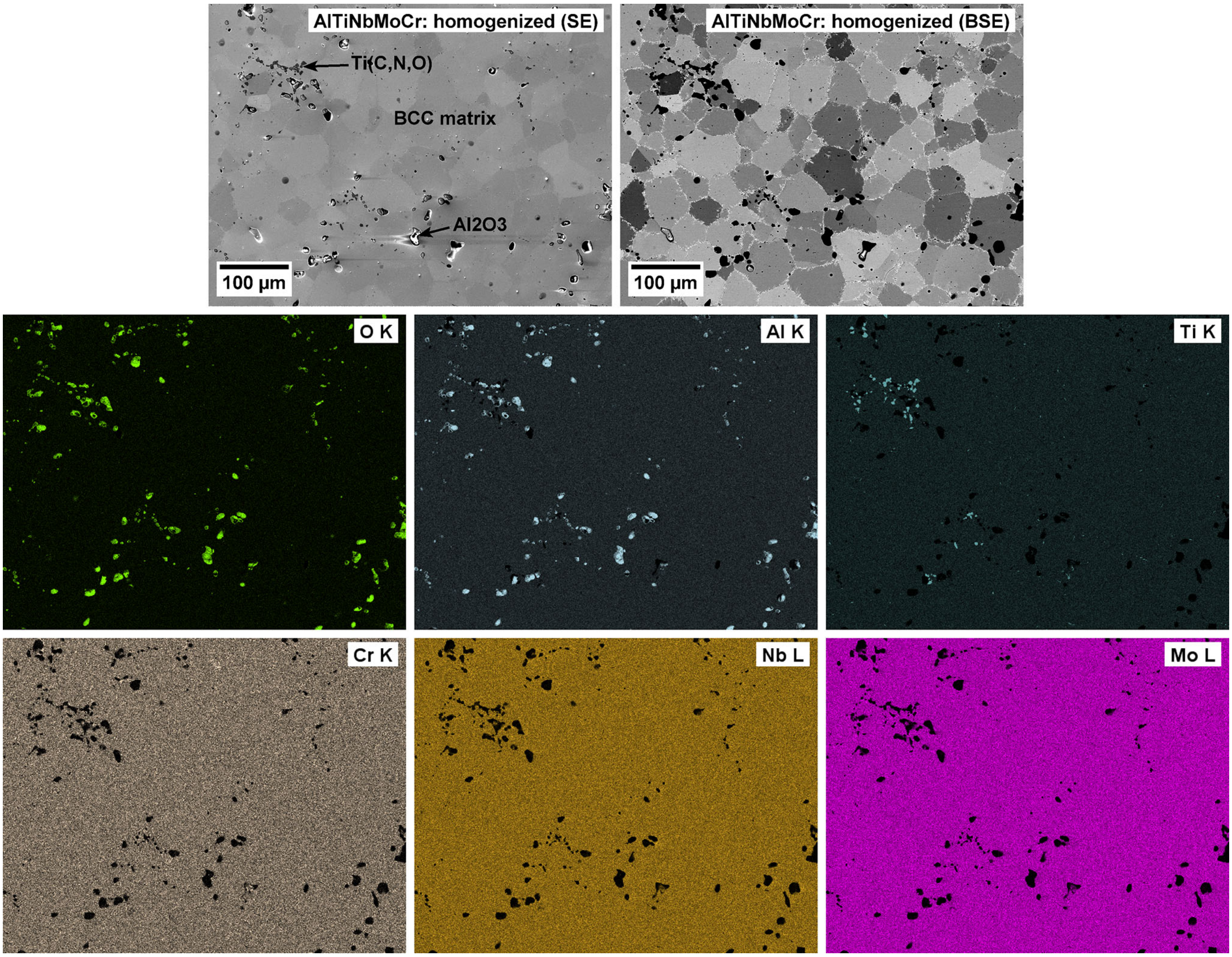

Fig. 11—Microstructure of the *homogenized* AlTiNb**MoCr** and corresponding EDS maps.

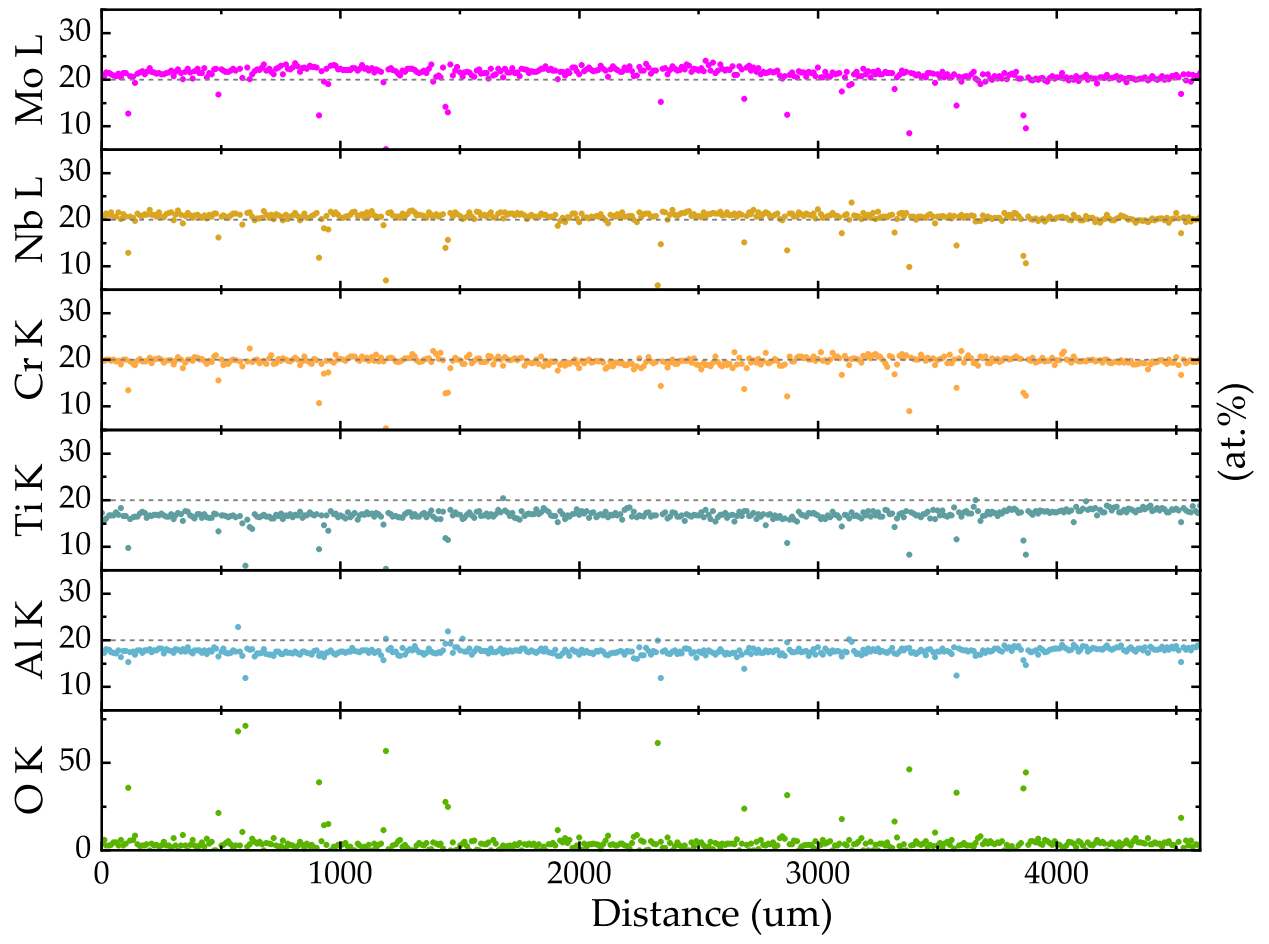

Fig. 12—EDS line scan of the *homogenized* AlTiNb**MoCr** over 4.6 mm showing chemical homogeneity at the scale of the former powder particles. The concentrations should be regarded as qualitative only, as the dwell time (and the number of counts) per point is too low for an accurate quantification. The outlying points correspond to the ceramic ($Al_2O_3$) particles.

On the other hand, the phase based on $Cr_3Si$ observed by XRD (Figure 14), was not predicted by CALPHAD at 1400 °C. However, when we plot the equilibrium phase fractions over the range of temperatures (Figure 15), we can see that this phase is predicted below 1050 °C. Again, CALPHAD correctly predicts that this phase may form in the given alloy composition.

The predicted decomposition of the BCC matrix in the AlTiNb**TaZr** alloy below 800 °C can also be observed in the Figure 15. However, it should be noted that such process may be strongly dependent on the chemical composition of the decomposing matrix. We know already that due to the absence of Nb and Ta in the AL3ZR5_D8M phase definition, the calculated composition of the matrix cannot be correct either, which may change the decomposition behavior.

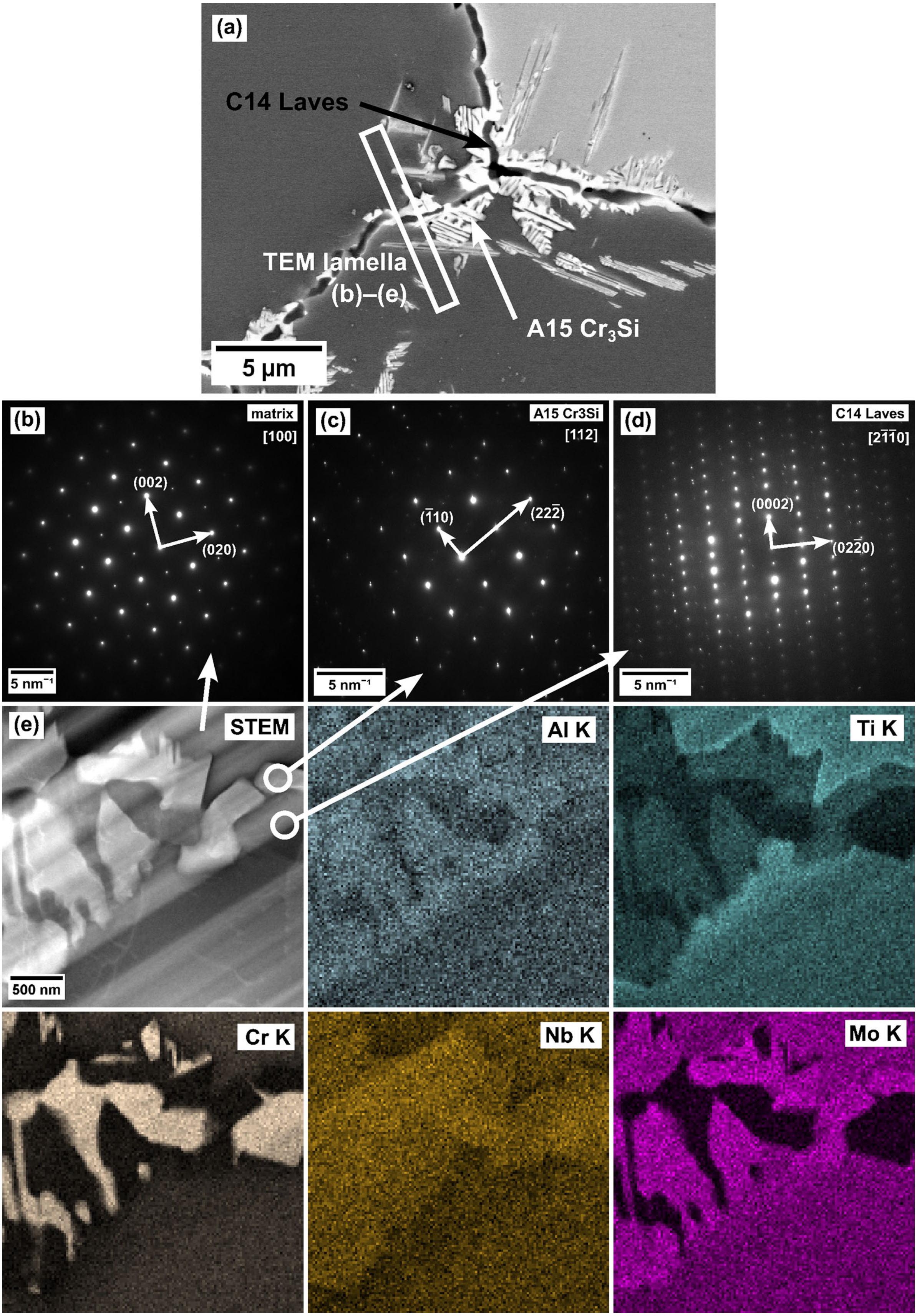

Fig. 13—A detail of the precipitates at the grain boundaries of the AlTiNb**MoCr** alloy. (*a*) BSE SEM micrograph of the area, (*b*) through (*d*) SAED patterns of individual phases, (e) STEM micrograph + EDS maps of the precipitates.

Finally, while the decomposition of AlTiNb**MoV** at temperatures below 1000 °C is also predicted, no such behavior was observed experimentally. This may be caused either by inaccuracy in the calculations, or by slow decomposition kinetics during cooling. Prediction for 1400 °C is, however, consistent with the experimental observations.

In conclusion, the CALPHAD calculations showed good qualitative agreement with the experimental results, although some disagreements were found. More importantly, CALPHAD offers more systematic approach to thermodynamic calculations, when compared with the semiempirical parameters (which often require some thermodynamic quantities as an input anyway). Our findings match those observed by Torralba *et al.* in Reference 30 where a satisfying agreement of experimental observations with CALPHAD calculations was reported.

However, it must be noted that the presence of light elements, namely N and O affects phase constitution. The lack of assessment of light elements, whose presence as the contaminants is inevitable, can be considered a significant shortcoming of the current databases. To our knowledge, in the most current TCHEA6 database, nitrogen was included, nevertheless, the oxygen influence on the phase equilibria should be considered as well. Presence of interstitial impurities can significantly change the phase stability similarly to Ti alloys.[24,31,32]

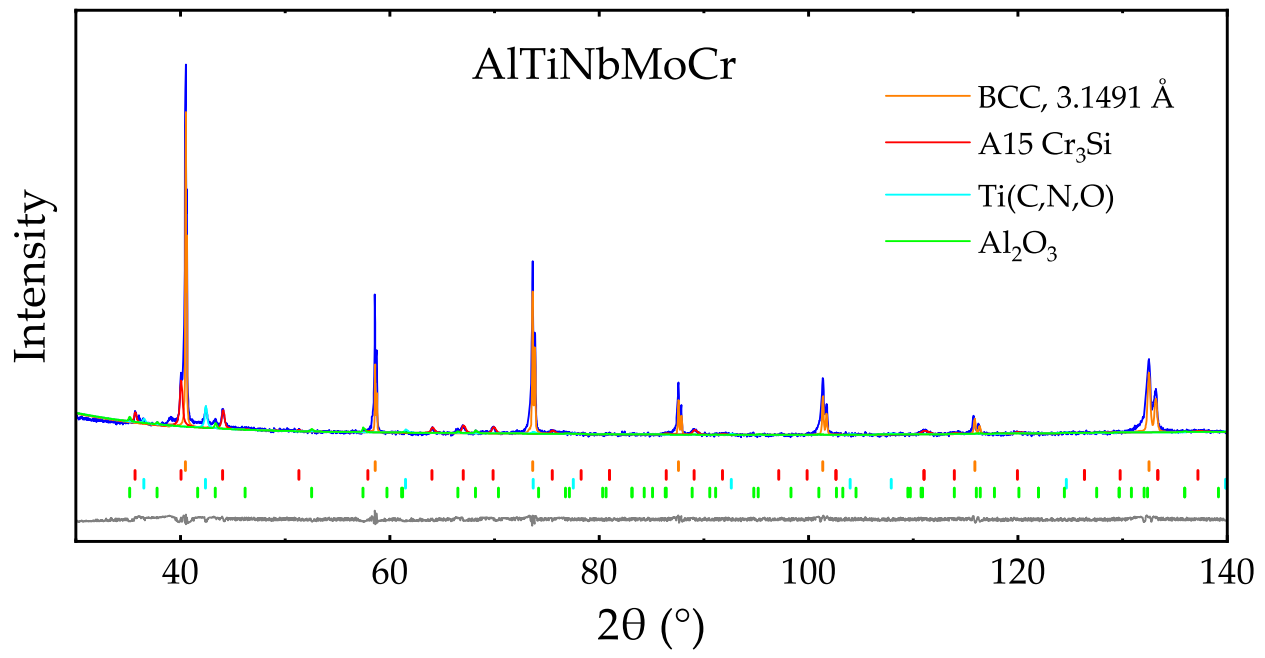

Fig. 14—X-ray diffractogram of the AlTiNb**MoCr** alloy. Four phases were identified: a dominant solid solution BCC phase and minor Ti(C,N,O), corundum and A15 $Cr_3Si$ phases.

## V. CONCLUSIONS

Phase composition of the three experimentally prepared RCCAs was compared to the predictions using semiempirical parameters and CALPHAD. The alloys were prepared from elemental powders by field-assisted sintering technology and successfully homogenized at 1400 °C after 168 hours.

- The homogenized AlTiNb**MoV** alloy exhibited a single-phase BCC microstructure with dispersed minor impurities ($Al_2O_3$ and Ti(C,N,O)). This is consistent with the CALPHAD predictions. No other phases have precipitated during cooling.
- The homogenized AlTiNb**TaZr** alloy exhibited a two-phase microstructure consisting of BCC matrix and $Al_3Zr_5$ based intermetallics. This is only qualitatively consistent with the CALPHAD predictions. Significant limitation of CALPHAD to predict compositionally complex intermetallic phases was revealed. Secondary BCC phase emerged as cuboidal precipitates within the matrix during cooling.
- Finally, the AlTiNb**MoCr** alloy exhibited a single-phase BCC microstructure at the homogenization temperature (with minor impurities), contrary to the CAPLHAD predictions of intermetallic Laves phase (C14). This phase precipitated during cooling, along with another phase, identified as $Cr_3Si$-based intermetallic phase (correctly predicted).

The accuracy of phase composition predictions via semiempirical parameters was found to be relatively low. Out of them, only the temperature insensitive $\Delta\chi + \delta r$ and $\Omega + \delta r$ parameters and $\phi(T_m)$ criterion performed reasonably, correctly predicting two of the three alloy studied. We conclude that predictions by CALPHAD are significantly more reliable than by semiempirical parameters and should be preferred for prediction of phase composition of RCCA.

**Table VI. Chemical and Phase Composition of the AlTiNbMoCr Alloy as Determined by EDS and XRD**

| | XRD | | EDS | | | | | | | |
|---|---|---|---|---|---|---|---|---|---|---|
| | Structure | Lattice Parameters (Å) | C (At. Pct) | N (At. Pct) | O (At. Pct) | Al (At. Pct) | Ti (At. Pct) | Nb (At. Pct) | Cr (At. Pct) | Mo (At. Pct) |
| Matrix | BCC | $a$ = 3.149 | — | — | — | 17.5 | 18.2 | 21.6 | 20.4 | 22.4 |
| A15 $Cr_3Si$ | $Pm\bar{3}n$ | $a$ = 5.0328 | — | — | — | 18* | 16* | 26* | 11* | 29* |
| C14 Laves | $P6_3/mmc$ | $a$ = 4.4* $c$ = 8.4* | — | — | — | 14* | 9* | 25* | 47* | 5* |
| $Al_2O_3$ | $R\bar{3}c$ | $a$ = 4.7609 $c$ = 13.007 | — | — | 66.5 | 31.5 | 1.0 | 0.4 | 0.4 | 0.4 |
| Ti(C,N,O) | FCC | $a$ = 4.2636 | 16.1 | 17.6 | 6.2 | 1.5 | 50.7 | 2.4 | 4.1 | 2.1 |

The Ti(C,N,O) chemical composition should be regarded as qualitative only, as the quantification of light elements overlapping with Ti is not precise. The lattice parameters and EDS marked with asterisk (*) were calculated from the TEM data.

**Table VII. Semi-empirical and Thermodynamic Parameters and Used for the Phase Predictions in the Investigated Refractory Complex Concentrated Alloys**

| Sample | e/a | $\delta r$ (Pct) | $\Delta\chi$ | $\overline{T_m}$ (K) | $\Delta H_{mix}$ kJ/mol | $\Delta S_{mix}$ J/(mol.K) | $\lvert\Delta S_E\rvert$ J/(mol.K) | $\Omega(\overline{T_m})$ | $\Omega(T_A)$ | $\phi(\overline{T_m})$ | $\phi(T_A)$ | $\Phi(\overline{T_m})$ | $\Phi(T_A)$ | $\kappa_2$ | $\kappa_1^{cr}(T)-\kappa_1$ |
|---|---|---|---|---|---|---|---|---|---|---|---|---|---|---|---|
| **AlTiNb**MoV | 1.8 | 3.79 | 5.60 | 2141 | − 12.8 | 13.38 | 0.315 | 2.24 | 1.75 | 23.52 | 18.20 | 0.33 | 0.28 | 0.6 | − 0.60 |
| **AlTiNb**TaZr | 2 | 4.55 | 7.35 | 2208 | − 16.16 | | 0.474 | 1.83 | 1.39 | 12.79 | 7.85 | 0.25 | 0.21 | 0.57 | 0.06 |
| **AlTiNb**MoCr | 1.6 | 5.48 | 7.16 | 2140 | − 12.64 | | 0.643 | 2.27 | 1.77 | 11.63 | 9.06 | 0.32 | 0.27 | 0.6 | − 0.43 |

The actual annealing temperature $T_A = 1673$K for all alloys

**Table VIII. Comparison of the Experimental Phase Composition and the Composition Predicted by Semiempirical Parameters**

| Sample | | Criterion Single Phase Condition | | | | | | | | |
|---|---|---|---|---|---|---|---|---|---|---|
| | | $\Omega + \delta r$[7] | | $\Delta H_{mix} + \delta r$[28] | $\Delta\chi + \delta r$[5] | $\phi$[10] | | $\Phi + \delta r$[12] | | $\kappa_1^{cr}(T)-\kappa_1$[14] |
| | | $\Omega \geq 1.1$ $\delta r < 6.6$ Pct | | $-11.6 < \Delta H_{mix} < 3.2$ $\delta r < 6.6$ Pct | $3 < \Delta\chi < 6$ Pct $\delta r < 6.6$ Pct | $\varphi > 20$ Pct | | $\Phi \geq 1$ $\delta r < 6.6$ Pct | | $\kappa_1^{cr} - \kappa_1 > 0$ |
| Temperature | Experimental Phase Composition at 1400 °C | $\overline{T_m}$ | $T_A$ | not used | not used | $\overline{T_m}$ | $T_A$ | $\overline{T_m}$ | $T_A$ | $T_A$ |
| **AlTiNb**MoV | SS | **SS** | **SS** | SS + IM | **SS** | **SS** | SS + IM | SS + IM | SS + IM | SS + IM |
| **AlTiNb**TaZr | SS + IM | SS | SS | **SS + IM** | **SS + IM** | **SS + IM** | **SS + IM** | **SS + IM** | **SS + IM** | SS |
| **AlTiNb**MoCr | SS | **SS** | **SS** | SS + IM | SS + IM | SS + IM | SS + IM | SS + IM | SS + IM | SS + IM |

For each criterion, the condition for single-phase solid solution is presented below. The entries in **bold** are in accordance with the experimental observations. The actual annealing temperature $T_A = 1673$K for all alloys.

**Table IX. Comparison of the CAPLHAD Prediction at 1400 °C (TCHEA4 Database, Rows in *italic*) and the Experimental Results**

| Phase | Al (At. Pct) | Ti (At. Pct) | Nb (At. Pct) | V (At. Pct) | Cr (At. Pct) | Zr (At. Pct) | Mo (At. Pct) | Ta (At. Pct) |
|---|---|---|---|---|---|---|---|---|
| **AlTiNb**MoV | | | | | | | | |
| BCC matrix | 18.1 | 19.2 | 20.5 | 18.9 | — | — | 23.3 | — |
| ***BCC_B2*** | *20* | *20* | *20* | *20* | — | — | *20* | — |
| **AlTiNb**TaZr | | | | | | | | |
| BCC_1 + BCC_2 matrix | 16.0 | 18.8 | 21.6 | — | — | 17.8 | — | 25.9 |
| $Al_3(Zr,Ti,Nb,Ta)_5$ | 33.9 | 10.8 | 8.8 | — | — | 39.6 | — | 6.9 |
| ***BCC_B2*** | *16.2* | *21.6* | *24.3* | — | — | *13.6* | — | *24.3* |
| ***AL3ZR5_D8M*** | *37.5* | *12.7* | *0* | — | — | *49.8* | — | *0* |
| **AlTiNb**MoCr | | | | | | | | |
| BCC/B2 matrix | 17.5 | 18.2 | 21.6 | — | 20.4 | — | 22.4 | — |
| C14 Laves | 14* | 9* | 25* | — | 47* | — | 5* | — |
| A15 $Cr_3Si$ | 18* | 16* | 26* | — | 11* | — | 29* | — |
| ***BCC_B2*** | *18.9* | *23.6* | *18.4* | — | *13.3* | — | *25.8* | — |
| ***C14_LAVES*** | *23.7* | *8.2* | *25.3* | — | *41.8* | — | *1.0* | — |

The starred values are from TEM EDS.

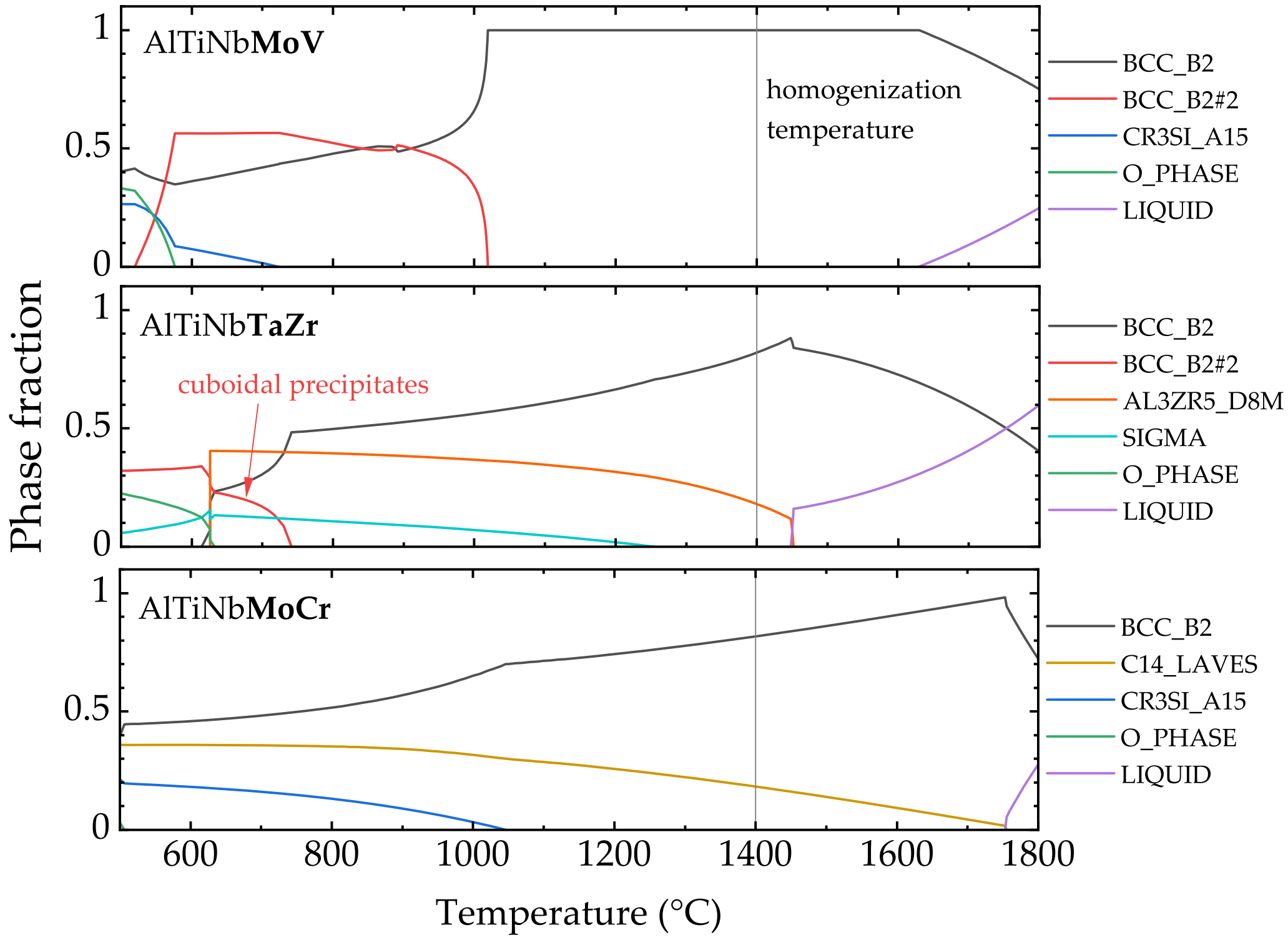

Fig. 15—Calculated equilibrium phase composition over the range of temperatures. The names of the phases were kept as in the TCHEA4 database definition.

## ACKNOWLEDGMENTS

This work was financially supported by Czech Science Foundation under the Project No. 22-24563S and by the Charles University Grant Agency, Project No. 292422. Financial support by the Operational Programme Johannes Amos Comenius of the MEYS of the Czech Republic, within the frame of the project Ferroic Multifunctionalities (FerrMion) [project No. CZ.02.01.01/00/22_008/0004591], co-funded by the European Union is also gratefully acknowledged.

## FUNDING

Open access publishing supported by the National Technical Library in Prague.

## DATA AVAILABILITY

The data that support the findings of this study are openly available in the Zenodo repository at https://doi.org/10.5281/zenodo.13834660 under the CC-BY 4.0 license.

## CONFLICT OF INTEREST

On behalf of all authors, the corresponding author states that there is no conflict of interest.